\documentclass[11pt,twoside]{article}
\usepackage{asp2014}

\aspSuppressVolSlug
\resetcounters

\begin{document}

\title{Fast interactive web-based data visualizer of panoramic spectroscopic surveys}

% full name: Ivan Katkov
\author{Ivan Katkov,$^{1,2}$ Damir Gasymov,$^{2,3}$ Joseph D. Gelfand,$^{1}$ Viktoria Toptun,$^{2,3}$ Kirill Grishin,$^{2}$ Igor Chilingarian$^{4,2}$ Anastasia Kasparova,$^{2}$ Vladislav Klochkov$^{2,3}$ Evgenii Rubtsov,$^{2,3}$ Vladimir Goradzhanov$^{2,3}$}
\affil{$^1$New York University Abu Dhabi, Abu Dhabi, United Arab Emirates; \email{ivan.katkov@nyu.edu}}
\affil{$^2$Sternberg Astronomical Institute, M.V.~Lomonosov Moscow State University, Moscow, Russia}
\affil{$^3$Department of Physics, M.V. Lomonosov Moscow State University, Moscow, Russia}
\affil{$^4$Smithsonian Astrophysical Observatory, Cambridge, MA, USA}

% remove/add as you need

% remove/add authors as you need
\paperauthor{Ivan~Katkov}{ivan.katkov@nyu.edu}{0000-0002-6425-6879}{New York University Abu Dhabi}{Center for Astro, Particle, and Planetary Physics}{Abu Dhabi}{Abu Dhabi}{129188}{United Arab Emirates}
\paperauthor{Damir~Gasymov}{gasymov.df18@physics.msu.ru}{0000-0002-1750-2096}{M.V.~Lomonosov Moscow State University}{Sternberg Astronomical Institute}{Moscow}{}{119992}{Russia}
\paperauthor{Joseph Gelfand}{joseph.gelfand@nyu.edu}{0000-0003-4679-1058}{New York University Abu Dhabi}{Center for Astro, Particle, and Planetary Physics}{Abu Dhabi}{Abu Dhabi}{129188}{United Arab Emirates}
\paperauthor{Viktoria~Toptun}{victoria.toptun@voxastro.org}{0000-0003-3599-3877}{M.V.~Lomonosov Moscow State University}{Sternberg Astronomical Institute}{Moscow}{}{119992}{Russia}
\paperauthor{Kirill~Grishin}{kirillg6@gmail.com}{0000-0003-3255-7340}{M.V.~Lomonosov Moscow State University}{Sternberg Astronomical Institute}{Moscow}{}{119992}{Russia}
\paperauthor{Igor Chilingarian}{igor.chilingarian@cfa.harvard.edu}{0000-0002-7924-3253}{Center for Astrophysics - Harvard and Smithsonian}{}{Cambridge}{}{02138}{USA}
\paperauthor{Anastasia~Kasparova}{anastasya.kasparova@gmail.com}{0000-0002-1091-5146}{M.V.~Lomonosov Moscow State University}{Sternberg Astronomical Institute}{Moscow}{}{119992}{Russia}
\paperauthor{Vladislav Klochkov}{vladislavk4481@gmail.com}{0000-0003-3095-8933}{M.V. Lomonosov Moscow State University}{Department of Physics}{Moscow}{}{119991}{Russia}
\paperauthor{Evgenii Rubtsov}{rubtsov602@gmail.com}{0000-0001-8427-0240}{Sternberg Astronomical Institute, Lomonosov Moscow State University}{}{Moscow}{}{119234}{Russia}
\paperauthor{Vladimir Goradzhanov}{goradzhanov.vs17@physics.msu.ru}{0000-0002-2550-2520}{Sternberg Astronomical Institute, Lomonosov Moscow State University}{}{Moscow}{}{119234}{Russia}

% remove/add as you need

% leave these next few aindex lines commented for the editors to enable them. Use Aindex.py to generate them for yourself.
% first presenting author should be the first entry for bold-facing the author index page-reference
%\aindex{Katkov,~I.}
%\aindex{Gasymov,~D.}
%\aindex{Gelfand,~J.~D.}
%\aindex{Grishin,~K.}
%\aindex{Toptun,~V.}
%\aindex{Kasparova,~A.}
%\aindex{Rubtsov,~E.}
%\aindex{Chilingarian,~I.}
% remove/add as you need

% leave the ssindex lines commented for the editors to enable them, use Index.py to suggest yours
%\ssindex{FOOBAR!conference!ADASS 2020}
%\ssindex{FOOBAR!organisations!ASP}

% leave the ooindex lines commented for the editors to enable them, use ascl.py to suggest yours
%\ooindex{FOOBAR, ascl:1101.010}
  
\begin{abstract}

Panoramic IFU spectroscopy is a core tool of modern observational astronomy and is especially important for galaxy physics. Many massive IFU surveys, such as SDSS MaNGA (10k targets), SAMI (3k targets), Califa (600 objects), Atlas3D (260 objects) have recently been released and made publicly available to the broad astronomical community. The complexity and massiveness of the derived data products from spectral cubes makes visualization of the entire dataset challenging, but nevertheless very important and crucial for scientific output.

Based on our past experience with visualization of spectral and imaging data built in the frame of the VOxAstro Initiative projects, we are now developing online web service for interactive visualizing spectroscopic IFU datasets (\url{ifu.voxastro.org}). Our service will provide a convenient access and visualization tool for spectral cubes from publicly available surveys (MaNGA, SAMI, Califa, Atlas3D) and results of their modeling, as well as maps of parameters derived from cubes, implementing the connected views concept. Here we describe the core components and functionality of the service, including REST API implementation on top of the Django+Postgres backend as well as a fast and responsive user interface built using the modern Vue.js-based framework Quasar.
  
\end{abstract}

\section{Introduction}

Massive spectroscopic surveys of galaxies provide extremely important information for studying galaxy evolution such as stellar and ionized gas kinematics, star formation history imprinted in the stellar population properties.
Integral Field Spectroscopy (IFS) is a powerful observational technique that allows spatially resolved measurements of galaxy properties.

Here we present an online web service (hereafter IFU Visualiser) for interactive visualization of spectroscopic IFU datasets (\url{https://ifu.voxastro.org}), providing a convenient access and visualization tool for spectral cubes from publicly available surveys (MaNGA, SAMI, Califa, Atlas3D).
The service is inspired by our first prototype \url{http://manga.voxastro.org} and the official Marvin web application released by the MaNGA Team \citep{Cherinka2019AJ....158...74C}.

\section{Data sets and service functionality}

To date (Nov 2021) IFU Visualiser involves data arrived from SDSS DR16 MaNGA Survey \citep{Bundy2015ApJ...798....7B}, SAMI DR3 \citep{Croom2012MNRAS.421..872C}, Califa DR3 \citep{Sanchez2012A&A...538A...8S}, Atlas3D \citep{Cappellari2011MNRAS.413..813C}. 
Table~\ref{table_xmatch} shows the size of the datasets available in the IFU Visualiser service and the cross-matching between surveys.

\begin{table}[!ht]
\caption{IFU Visualiser data sets}
\label{table_xmatch}
\smallskip
\begin{center}
{\small
\begin{tabular}{llllll}  % l = left, c = centered
\tableline
\noalign{\smallskip}
 & MaNGA DR17 & MaNGA DR16 & SAMI & Califa & Atlas3D \\
\noalign{\smallskip}
\tableline
\noalign{\smallskip}
MaNGA DR17  & \textbf{11236} & 4824         & 106           & 40            & 0 \\
MaNGA DR16  &        4824   & \textbf{4824} & 15            & 20            & 0 \\
SAMI        &        106    & 15            & \textbf{3426} & 1             & 1 \\
Califa      &        40     & 20            & 1             & \textbf{667}  & 23 \\
Atlas3D     &        0      &  0            & 1             & 23            & 260 \\
\noalign{\smallskip}
\tableline
\noalign{\smallskip}
\end{tabular}
}
\end{center}
\end{table}

IFU Visualiser largely inherits the general architecture and functionality of another VOxAstro project RCSEDv2 \citep{Chilingarian2017ApJS..228...14C,O7-009_Toptun_RCSEDv2,O3-006_Klochkov_RCSEDv2}.
Briefly, IFU Visualiser provides two major ingredients: search tool for spectral cubes and interactive visualiser of individual cubes and two-dimensional maps of parameters extracted from spectra. Query language syntax described on the documentation page https://ifu.voxastro.org/docs where several search query examples can be found. Fig.~\ref{fig1_sp_viewer} shows a screenshot of the individual Cube page.

\articlefigure{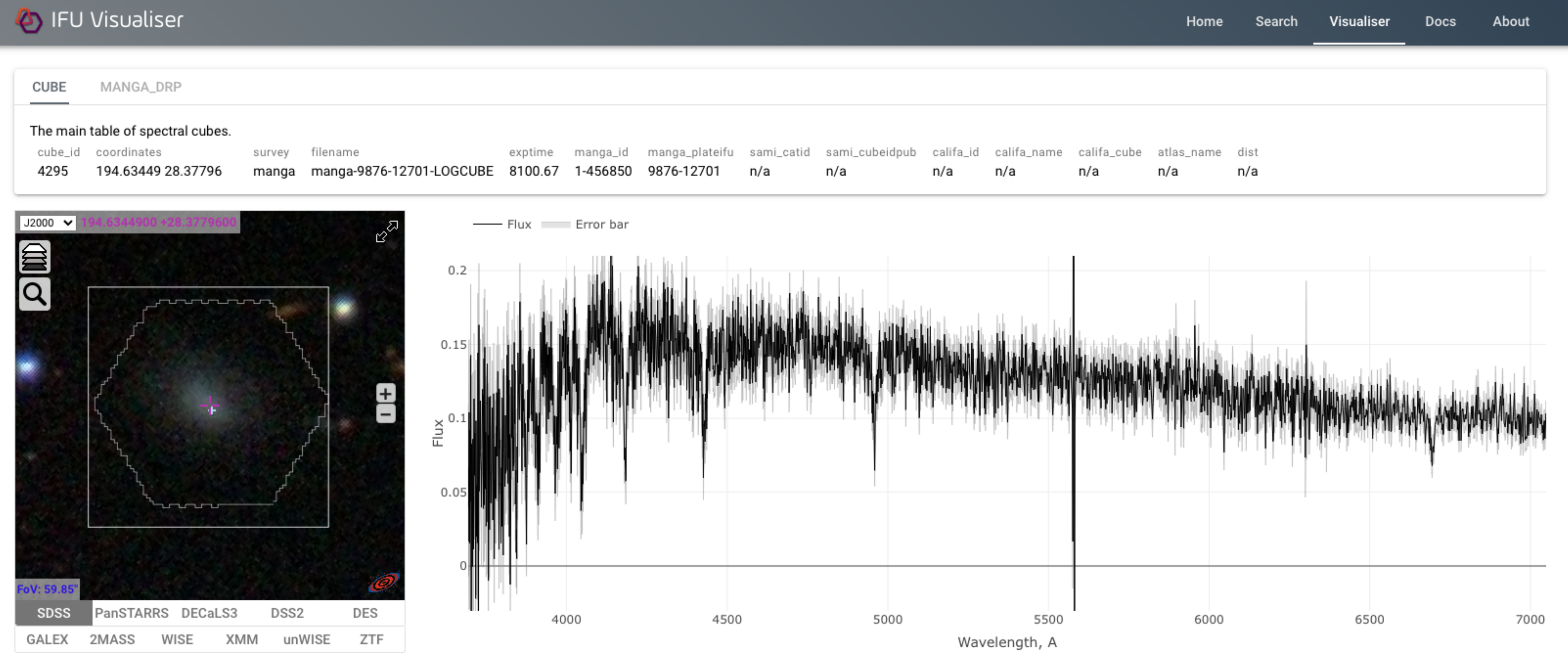}{fig1_sp_viewer}{
Example of an individual page of the MaNGA spectral cube (Cube ID: 4295) showing 
the low-mass post-starburst galaxy GMP~4188 in the Coma cluster, which will passively evolve into an ultra-diffuse galaxy \citep{Grishin2021NatAs.tmp..208G}.
}

\section{Implementation}

All code, including the backend, frontend, and scripts to assembly the input tables for the database, is available on the GitHub repo.\footnote{\url{https://github.com/voxastro/ifu-visualiser}}

\subsection{Backend}

PostgreSQL is used to manage access to the main and linked tables.
The main spectral cube table contains basic information such as cube identifier, coordinates, field-of-view coverage, and the necessary survey identifiers needed to connect related tables to any given cube.
The linked tables are taken from the corresponding survey and contain very diverse information about cubes as well as galaxies. 
We chose PostgreSQL as the backend database because of i) support for geometry queries thanks to q3c \citep{Koposov2006ASPC..351..735K} and pgSphere \citep{Chilingarian2004ASPC..314..225C} extensions; and ii) natural compatibility with VO services (TAP, SSAP access) using GAVO DaCHS \citep{Demleitner2018ascl.soft04005D}, which we plan to add soon to IFU Visualiser.
The well-known Python-based Django REST Framework (DRF)\footnote{\url{https://www.django-rest-framework.org/}} is used on top of the PostgreSQL database to provide API endpoints.
In addition to Django high popularity, wide community, and excellent documentation, a decisive reason for using this framework is the large number of batteries.
One of them is \verb"django-custom-query",\footnote{\url{https://github.com/ivan-katkov/django-custom-query}} which is designed to create simplified, but still very powerful SQL-like queries (e.g. \verb|survey=sami OR atlas_name~4551|). We also modified the original library and incorporated coordinates Cone-search queries using the q3c function \verb"q3c_radial_query()" (e.g. \verb"cone(195, 28, 0.5)").
We also considered using the GraphQL\footnote{\url{https://graphql.org/}} endpoint to provide data access, but after simple performance tests with Graphene-Django library\footnote{\url{https://docs.graphene-python.org/projects/django/en/latest/}} we found that DRF provides several times faster response time overall.
Another problem with using GraphQL lies on the frontend side: the seeming simplicity of using, for example, with Apollo client\footnote{\url{https://apollo.vuejs.org/}} preserves potential problems in managing global application state.
The Apollo client can be used as a state manager, but we use the standard Vuex state manager for Vue.js applications.
Using both options leads to the "two sources of truth" problem. Switching completely to Apollo requires writing awkward chunks of code and abandoning browser testing tools not yet available for the Apollo client.

\subsection{Frontend}

We used the Vue.js-based frontend framework Quasar\footnote{\url{https://quasar.dev/}}, which provides an awesome infrastructure and a large collection of Vue components built according to Material Design guidelines. Vue.js and the Quasar framework have a very low threshold of entry and allow one to create highly interactive user interfaces with very limited Javascript skills.
In addition to the Quasar component library, we used Aladin Lite for imaging and Plotly.js\footnote{\url{https://plotly.com/javascript/}} library to create an interactive spectral plots and maps of parameters.

\section{Next development steps}

To date (November 2021), only a small part of the planned functionality of the IFU Visualiser has been developed.
The next major steps for further development will be: 
\begin{enumerate}
    \item Regarding user interface development we will improve the spectral Cube page and implementing a more flexible interaction with spectra and parameter maps.
    Functionality will be similar to that we built in the frame of our first prototype of such a service \url{https://manga.voxastro.org/} made with another  technologies and uses only MaNGA data.
    \item We also plan to add uniform basic galaxy parameters such as masses, luminosities, effective radii and colors for all galaxies  available in the IFU Visualiser.
    \item Another important milestone is to analyze the entire collection of spectral cubes to ensure a homogeneous set of data products extracted from the cubes using full spectral fitting technique NBursts, similar to what we do in the RCSED2 project but for IFU data.
\end{enumerate}

\textbf{Acknowledgments.}
This project is supported by the Russian Science Foundation grant 21-72-00036. Authors also thank the Interdisciplinary Scientific and Educational School of Moscow University ``Fundamental and Applied Space Research''.
DG, KG, VT, EG are also thankful to the ADASS XXXI organizing committee for
providing financial aid.

\bibliography{O3-005}

\end{document}